# Class-based Rough Approximation with Dominance Principle

**Junyi Chai, James N.K. Liu**
The Hong Kong Polytechnic University
E-mail: {csjchai,csnkliu}@comp.polyu.edu.hk

## Abstract

*Dominance-based Rough Set Approach (DRSA), as the extension of Pawlak's Rough Set theory, is effective and fundamentally important in Multiple Criteria Decision Analysis (MCDA). In previous DRSA models, the definitions of the upper and lower approximations are preserving the class unions rather than the singleton class. In this paper, we propose a new Class-based Rough Approximation with respect to a series of previous DRSA models, including Classical DRSA model, VC-DRSA model and VP-DRSA model. In addition, the new class-based reducts are investigated.*

## 1. Introduction

Multiple Criteria Decision Analysis (MCDA) aims at providing the decision maker (DM) a knowledge recommendation while considering the finite objects evaluated from multiple viewpoints (known as *criteria*). Roy [9] considered four problems in MCDA, including *criteria analysis*, *choice*, *ranking*, *sorting*. The first one is the essential procedure for optimization of decision information and the latter three ones can produce specific decision outcomes.

Apart from several valid and classical MCDA approaches (see the state-of-the-art survey in [3]), the non-classical methods and techniques (like [1][2]) are significant since it attempts to address the risk and uncertainty of MCDA catering to the real world. Classical Rough Set Approach (CRSA for short) initially proposed by Pawlak (see [8]) is an effective mathematical tool for decision analysis. But, it fails to deal with the preference-ordered data in MCDA. In this reason, Dominance-based Rough Set Approach (DRSA for short) was generated by Greco and his colleagues [5][10]. Unlike the CRSA which makes use of the indiscernibility relations for construction of knowledge granular, DRSA considers the dominance relations of these preference-ordered data in given decision table.

The target by using DRSA is to induce the decision rule as classifier for providing the suitable assignment of both learning objects (from given decision table) and new objects. Recently, the classical DRSA model had been extended to VC-DRSA [4], VP-DRSA [6], etc.

In all previous DRSA models, the upper and lower approximations are defined in consideration of the union of decision class (i.e. upward union $Cl_t^{\geq}$ and downward union $Cl_t^{\leq}$). We call them as *union-based rough approximation*. In this paper, we attempt to investigate the issue: whether one singleton decision class can be used to define the upper and lower approximation in a series of DRSA models. To this end, we firstly analyze the partition of objects preserving one particular decision class, and provide a new Three Region Model (TRM). Then, we develop the so-called *class-based rough approximation* in a series of previous DRSA models, including the classical DRSA model, VC-DRSA model and VP-DRSA model. Finally, inspired by Inuiguchi's initial works [6][7], the class-based criteria reduction is also studied.

This paper is organized as follows: The next section briefly reviews the basic principles of DRSA theory. Section 3 studies the class-based rough approximation in a series of DRSA models. Section 4 investigates the class-based criteria reduction. Finally, we draw the conclusion in section 5.

## 2. Background

In this section, we concisely revisit the basic theory of DRSA. Despite the various problem domains regarding MCDA, three elementary factors are usually involved, including objects, criteria and DM(s). These factors can generally be organized as *decision table* with columns of criteria and rows of objects. Formally, a decision table is the 4-tuple $S = \langle U, Q, V, f \rangle$, which includes (1) a finite set of objects denoted by $U$, $x \in U = \{x_1, ..., x_m\}$; (2) a finite set of criteria is denoted by

$Q = C \cup D$, where condition criteria set $C \neq \varnothing$, decision criteria set $D \neq \varnothing$ (usually the singleton set $D = \{d\}$), and $q \in Q = \{q_1, ..., q_n\}$; (3) the domain of criterion $q$ denoted by $V_q$, where $V = \cup_{q \in Q} V_q$; (4) information function denoted by $f_q(x): U \times Q \to V$, where $f_q(x) \in V_q$ for each $q \in Q$, $x \in U$.

The objective sets of rough approximations are the upward or downward unions of predefined decision classes (we call *union-based rough approximations*). Suppose the decision criterion $\{d\}$ makes a partition of $U$ into a finite number of classes $CL = \{Cl_t, t = 1,...,l\}$. We assume that $Cl_{t+1}$ is superior to $Cl_t$ according to DM's preference. Each object $x \in U$ belongs to *one and only one* class $Cl_t \in CL$. The upward and downward unions of classes are represented as:

$$Cl_t^{\geq} = \bigcup_{s \geq t} Cl_s, \quad Cl_t^{\leq} = \bigcup_{s \leq t} Cl_s, \text{ where } t = 1,...,l.$$

Then, the following operational laws are valid:

$Cl_1^{\leq} = Cl_1$; $Cl_l^{\geq} = Cl_l$; $Cl_t^{\geq} = U - Cl_{t-1}^{\leq}$; $Cl_t^{\leq} = U - Cl_{t+1}^{\geq}$;

$Cl_1^{\geq} = Cl_l^{\leq} = CL$; $Cl_0^{\leq} = Cl_{l+1}^{\geq} = \varnothing$.

The knowledge granules in DRSA theory are *dominance cones*. If two decision values are with the dominance relation like $f_q(x) \geq f_q(y)$ for every considered criterion $q \in P \subseteq C$, we say $x$ *dominates* $y$, denoted by $x D_P y$. The dominance relation is reflexive and transitive. With this in mind, the *dominance cone* of object $x$ can be represented by:

$D_P^+(x) = \{y \in U : y D_P x\}$; $D_P^-(x) = \{y \in U : x D_P y\}$.

The key concept in DRSA theory is the *Dominance Principle*: if the decision value of object $x$ is no worse than that of object $y$ on all considered condition criteria (saying $x$ is dominating $y$ on $P \subseteq C$), object $x$ should also be assigned to a decision class no worse than that of object $y$ (saying $x$ is dominating $y$ on $D$). Objects satisfying the dominance principle are called *consistent*, and also, objects violating the dominance principle are called *inconsistent*. A decision table which contains *inconsistent object* is called *inconsistency table*. According to such dominance principle, the definition of rough approximations is given as follows.

P-lower approximation of class union $Cl_t^{\geq}$ and $Cl_t^{\leq}$, denoted by $\underline{P}(Cl_t^{\geq})$ and $\underline{P}(Cl_t^{\leq})$ respectively, are represented as:

$\underline{P}(Cl_t^{\geq}) = \{x \in U : D_P^+(x) \subseteq Cl_t^{\geq}\}$; $\underline{P}(Cl_t^{\leq}) = \{x \in U : D_P^-(x) \subseteq Cl_t^{\leq}\}$.

P-upper approximation of class union $Cl_t^{\geq}$ and $Cl_t^{\leq}$, denoted by $\overline{P}(Cl_t^{\geq})$ and $\overline{P}(Cl_t^{\leq})$ respectively, are represented as:

$\overline{P}(Cl_t^{\geq}) = \{x \in U : D_P^-(x) \cap Cl_t^{\geq} \neq \varnothing\}$;

$\overline{P}(Cl_t^{\leq}) = \{x \in U : D_P^+(x) \cap Cl_t^{\leq} \neq \varnothing\}$.

Rough boundary region of class union $Cl_t^{\geq}$ and $Cl_t^{\leq}$, denoted by $Bn_P(Cl_t^{\geq})$ and $Bn_P(Cl_t^{\leq})$ respectively, are represented as:

$Bn_P(Cl_t^{\geq}) = \overline{P}(Cl_t^{\geq}) - \underline{P}(Cl_t^{\geq})$; $Bn_P(Cl_t^{\leq}) = \overline{P}(Cl_t^{\leq}) - \underline{P}(Cl_t^{\leq})$.

Obviously, we have the properties:

$Bn_P(Cl_t^{\geq}) = Bn_P(Cl_{t-1}^{\leq}) = \overline{P}(Cl_t^{\geq}) \cap \overline{P}(Cl_{t-1}^{\leq})$.

In addition, the following properties are valid:

$\underline{P}(Cl_t^{\geq}) \subseteq Cl_t^{\geq} \subseteq \overline{P}(Cl_t^{\geq})$; $\underline{P}(Cl_t^{\leq}) \subseteq Cl_t^{\leq} \subseteq \overline{P}(Cl_t^{\leq})$;

$\underline{P}(Cl_t^{\geq}) = U - \overline{P}(Cl_{t-1}^{\leq})$; $\underline{P}(Cl_t^{\leq}) = U - \overline{P}(Cl_{t+1}^{\geq})$;

$\overline{P}(Cl_t^{\geq}) = U - \underline{P}(Cl_{t-1}^{\leq})$; $\overline{P}(Cl_t^{\leq}) = U - \underline{P}(Cl_{t+1}^{\geq})$.

If $Q \subseteq P \subseteq C$, we have the following properties:

$\underline{Q}(Cl_t^{\geq}) \subseteq \underline{P}(Cl_t^{\geq})$; $\overline{Q}(Cl_t^{\geq}) \supseteq \overline{P}(Cl_t^{\geq})$;

$\underline{Q}(Cl_t^{\leq}) \subseteq \underline{P}(Cl_t^{\leq})$; $\overline{Q}(Cl_t^{\leq}) \supseteq \overline{P}(Cl_t^{\leq})$.

The definitions of the classical DRSA model are based on the strict dominance principle (as shown in above). Inspired by the Variable Precision Rough Set [11], which is the extension of CRSA via relaxation of strict indiscernibility relation, Greco et al. [10] provided the VC-DRSA model. This model accepts a limited number of inconsistency objects controlled by a predefined threshold called consistency level.

The lower approximations of VC-DRSA model can be represented as follows. For any $P \subseteq C$, we have:

$\underline{P}^l(Cl_t^{\geq}) = \{x \in Cl_t^{\geq} : \frac{|D_P^+(x) \cap Cl_t^{\geq}|}{|D_P^+(x)|} \geq l\}$;

$\underline{P}^l(Cl_t^{\leq}) = \{x \in Cl_t^{\leq} : \frac{|D_P^-(x) \cap Cl_t^{\leq}|}{|D_P^-(x)|} \geq l\}$.

where $l$ is called consistency level, which means that object $x \in U$ belongs to $Cl_t^{\geq}$ (or $Cl_t^{\leq}$) with no ambiguity at level $l \in (0,1]$.

Based on the definitions of lower approximation, we can further obtain the definitions of the upper approximations and the rough boundary regions as:

$\overline{P}^l(Cl_t^{\geq}) = U - \underline{P}^l(Cl_{t-1}^{\leq})$; $Bn_P(Cl_t^{\geq}) = \overline{P}^l(Cl_t^{\geq}) - \underline{P}^l(Cl_t^{\geq})$;

$\overline{P}^l(Cl_t^{\leq}) = U - \underline{P}^l(Cl_{t+1}^{\geq})$; $Bn_P(Cl_t^{\leq}) = \overline{P}^l(Cl_t^{\leq}) - \underline{P}^l(Cl_t^{\leq})$.

## 3. Class-based rough approximation

### 3.1. Class-based classical DRSA model

Classical DRSA model can be regarded as a special case of VC-DRSA model with the consistency level fulfilling $l_1 = l_2 = 1$ (the strict dominance principle), while, $\frac{|D_P^+(x) \cap Cl_t^{\geq}|}{|D_P^+(x)|} \geq l_1$ and $\frac{|D_P^-(x) \cap Cl_t^{\leq}|}{|D_P^-(x)|} \geq l_2$.

Table 1. Constraint conditions of objects preserving decision class $Cl_t$.

| Classical DRSA model with $l_1 = l_2 = 1$ | Constraint Conditions | |
|---|---|---|
| | Low region | High region |
| $l_1$ in $Cl_{t-1}$ | (C') $\{x \in U : D_P^-(x) \subseteq Cl_{t-1}^{\leq}\}$ | (D') $\{x \in U : D_P^-(x) \cap Cl_t^{\geq} \neq \varnothing\}$ |
| $l_2$ in $Cl_t$ | (A) $\{x \in U : D_P^+(x) \cap Cl_{t-1}^{\leq} \neq \varnothing\}$ | (B) $\{x \in U : D_P^+(x) \subseteq Cl_t^{\geq}\}$ |
| $l_1$ in $Cl_t$ | (C) $\{x \in U : D_P^-(x) \subseteq Cl_t^{\leq}\}$ | (D) $\{x \in U : D_P^-(x) \cap Cl_{t+1}^{\geq} \neq \varnothing\}$ |
| $l_2$ in $Cl_{t+1}$ | (A') $\{x \in U : D_P^+(x) \cap Cl_t^{\leq} \neq \varnothing\}$ | (B') $\{x \in U : D_P^+(x) \subseteq Cl_{t+1}^{\geq}\}$ |

Table 2. Four regions model preserving object $x \in Cl_t$.

| objects: $x \in Cl_t$ | Fulfilled constraint condition in class $Cl_t$ | |
|---|---|---|
| | Consideration of $D_P^+(x)$ | Consideration of $D_P^-(x)$ |
| Region I: | (A): $\{x \in U : D_P^+(x) \cap Cl_{t-1}^{\leq} \neq \varnothing\}$ | (C): $\{x \in U : D_P^-(x) \subseteq Cl_t^{\leq}\}$ |
| Region II: | (B): $\{x \in U : D_P^+(x) \subseteq Cl_t^{\geq}\}$ | (C): $\{x \in U : D_P^-(x) \subseteq Cl_t^{\leq}\}$ |
| Region III: | (B): $\{x \in U : D_P^+(x) \subseteq Cl_t^{\geq}\}$ | (D): $\{x \in U : D_P^-(x) \cap Cl_{t+1}^{\geq} \neq \varnothing\}$ |
| Region IV: | (A): $\{x \in U : D_P^+(x) \cap Cl_{t-1}^{\leq} \neq \varnothing\}$ | (D): $\{x \in U : D_P^-(x) \cap Cl_{t+1}^{\geq} \neq \varnothing\}$ |

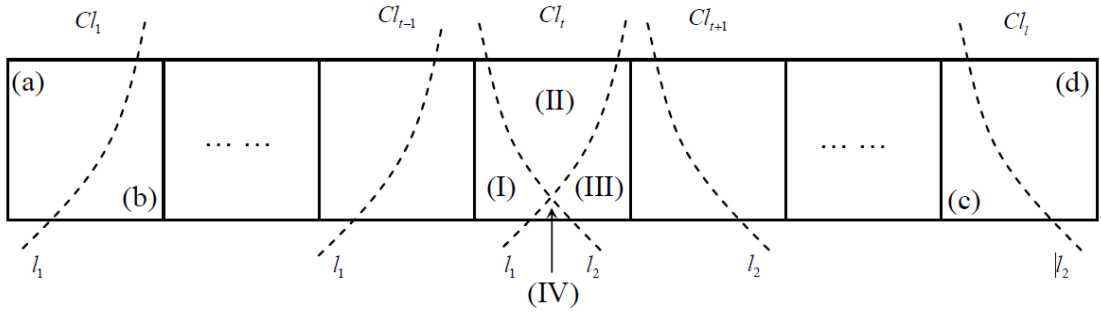

Figure 1. The decision class $CL$ as the partition of $U$ in DRSA models

With this in mind, we can obtain the constraint conditions from the definitions of rough approximation. These conditions are used to partition the objects which have been assigned to a singleton decision class. Considering the decision class $Cl_t$ and its two adjacent classes $Cl_{t-1}$ and $Cl_{t+1}$, the constraint conditions are given in Table 1. Each consistency level $l_1$ (or $l_2$) divides the entire objects into two regions: Low region and High region. These regions are constrained by different conditions. For the class $Cl_t$, the constraint conditions are (A), (B), (C), (D). For its adjacent classes $Cl_{t-1}$ and $Cl_{t+1}$, the constraint conditions are (A'), (B'), (C'), (D').

Considering the class $Cl_t$, object $x \in Cl_t$ must be assigned to one of the four regions, as Region I, II, III, IV. Each region is constrained by two conditions which are defined by the dominance cores: $D_P^+(x)$ and $D_P^-(x)$, respectively. The details are shown in Table 2. And,

Fig. 1 illustrates the partition of $U$ at consistency levels $l_1$ and $l_2$ in all decision classes $CL = \{Cl_t, t = 1, ..., l\}$.

Based on such observations, we consider three regions in class-based classical DRSA model with respect to the predefined $Cl_t$ ($t = 2, ..., l-1$):

- Low boundary region, denoted by $P_\beta(Cl_t)$:

  $P_\beta(Cl_t) = \{x \in Cl_t : D_P^+(x) \cap Cl_{t-1}^{\leq} \neq \varnothing\}$

- Precise classification region, denoted by $\underline{P}(Cl_t)$:

  $\underline{P}(Cl_t) = \{x \in Cl_t : D_P^+(x) \subseteq Cl_t^{\geq} \text{ and } D_P^-(x) \subseteq Cl_t^{\leq}\}$

- High boundary region, denoted by $P^\beta(Cl_t)$:

  $P^\beta(Cl_t) = \{x \in Cl_t : D_P^-(x) \cap Cl_{t+1}^{\geq} \neq \varnothing\}$

Particularly, there are just $\underline{P}(Cl_1)$ and $P^\beta(Cl_1)$ for class $Cl_1$ and just $P_\beta(Cl_l)$ and $\underline{P}(Cl_l)$ for class $Cl_l$.

Corresponding to the Fig. 1, we have the following assertions:

- Region $P_\beta(Cl_t)$ consists of Region I and IV.
- Region $\underline{P}(Cl_t)$ consists of Region II.

● Region $P^\beta(Cl_t)$ consists of Region III and IV.

We call the above definitions as Three Region Model (TRM).

Furthermore, if there are only two decision classes in a given decision table (i.e. Pairwise comparison table, profit or non-profit, right or wrong), the TRM can be represented as follows:

**Definition** (TRM in two-grade decision table)

According to the given decision table, the decision criteria $\{d\}$ makes a partition of $U$ into two classes $S$ and $S^c$ (suppose $S$ is superior to $S^c$ in DM's preference). And each $x \in U$ belongs to one and only one of such two predefined classes. The two-grade class-based rough approximations are represented as follows.

- Precise classification region of class $S$:
  $\underline{P}(S) = \{x \in S : D_P^+(x) \subseteq S\}$
- Low boundary region of class $S$:
  $P_\beta(S) = \{x \in S : D_P^+(x) \cap S^c \neq \emptyset\}$
- High boundary region of class $S^c$:
  $P^\beta(S^c) = \{x \in S^c : D_P^-(x) \cap S \neq \emptyset\}$
- Precise classification region of class $S^c$:
  $\underline{P}(S^c) = \{x \in S^c : D_P^-(x) \subseteq S^c\}$

And also, we can obtain following properties, which can be easily proved:

$P_\beta(S) = S - \underline{P}(S)$ ; $P^\beta(S^c) = S^c - \underline{P}(S^c)$ .

Next, we investigate the relationship between the definitions of the union-based and the class-based rough approximations. Considering the decision class $Cl_t \in CL$ ($t = 2,...,l-1$) and its adjacent classes $Cl_{t-1}$ and $Cl_{t+1}$, the following properties are valid:

$Cl_t = \underline{P}(Cl_t) + P_\beta(Cl_t) + P^\beta(Cl_t) - P_\beta(Cl_t) \cap P^\beta(Cl_t)$

$\underline{P}(Cl_t^\geq) \cap \underline{P}(Cl_t^\leq) = \{x \in Cl_t : D_P^+(x) \subseteq Cl_t^\geq \text{ and } D_P^-(x) \subseteq Cl_t^\leq\} = \underline{P}(Cl_t)$

$\underline{P}(Cl_t^\leq) \cap Cl_t = \{x_i \in Cl_t : D_P^-(x_i) \subseteq Cl_t^\leq\} = Cl_t - P^\beta(Cl_t)$

$\underline{P}(Cl_t^\geq) \cap Cl_t = \{x_i \in Cl_t : D_P^+(x_i) \subseteq Cl_t^\geq\} = Cl_t - P_\beta(Cl_t)$

$Bn_P(Cl_t^\geq) \cap Cl_t = \{x \in Cl_t : D_P^+(x) \cap Cl_{t-1}^\leq \neq \emptyset\} = P_\beta(Cl_t)$

$Bn_P(Cl_t^\leq) \cap Cl_t = \{x \in Cl_t : D_P^-(x) \cap Cl_{t+1}^\geq \neq \emptyset\} = P^\beta(Cl_t)$

$Bn_P(Cl_{t-1}^\leq) \cap Cl_{t-1} = Bn_P(Cl_t^\geq) \cap Cl_{t-1} = P^\beta(Cl_{t-1})$

$Bn_P(Cl_{t+1}^\geq) \cap Cl_{t+1} = Bn_P(Cl_t^\leq) \cap Cl_{t+1} = P_\beta(Cl_{t+1})$

$Bn_P(Cl_{t-1}^\leq) \supseteq (P^\beta(Cl_{t-1}) + P_\beta(Cl_t)) \subseteq Bn_P(Cl_t^\geq)$

$Bn_P(Cl_t^\leq) \supseteq (P^\beta(Cl_t) + P_\beta(Cl_{t+1})) \subseteq Bn_P(Cl_{t+1}^\geq)$

$\bigcup_{s \geq t}(Cl_s - P_\beta(Cl_s)) \subseteq \underline{P}(Cl_t^\geq)$

$\bigcup_{s \leq t}(Cl_s - P^\beta(Cl_s)) \subseteq \underline{P}(Cl_t^\leq)$

### 3.2. Class-based VC-DRSA model

In this section, we investigate the TRM in VC-DRSA model. For any $P \subseteq C$, we say that $x \in U$ belongs to $Cl_t^\geq$ at consistency level $l_2 \in (0,1]$, and $x \in U$ belongs to $Cl_t^\leq$ at consistency level $l_1 \in (0,1]$. The concept of lower approximations at some consistency levels $l_1$ and $l_2$ are formally presented as:

$\underline{P}^{l_2}(Cl_t^\geq) = \{x \in Cl_t^\geq : \frac{|D_P^+(x) \cap Cl_t^\geq|}{|D_P^+(x)|} \geq l_2\}$ , $t = 1,...,l$ ;

$\underline{P}^{l_1}(Cl_t^\leq) = \{x \in Cl_t^\leq : \frac{|D_P^-(x) \cap Cl_t^\leq|}{|D_P^-(x)|} \geq l_1\}$ , $t = 1,...,l$ .

Then, the TRM preserving the predefined class $Cl_t$ ($t = 2,...,l-1$) can be presented as:

● Low boundary region $P_\beta^{l_2}(Cl_t)$ :

$P_\beta^{l_2}(Cl_t) = Bn_P^{l_1 l_2}(Cl_t^\geq) \cap Cl_t = \{x \in Cl_t : \frac{|D_P^+(x) \cap Cl_t^\geq|}{|D_P^+(x)|} < l_2\}$

● Precision classification region $\underline{P}^{l_1 l_2}(Cl_t)$ :

$\underline{P}^{l_1 l_2}(Cl_t) = \underline{P}^{l_1}(Cl_t^\leq) \cap \underline{P}^{l_2}(Cl_t^\geq)$

$= \{x \in Cl_t : \frac{|D_P^-(x) \cap Cl_t^\leq|}{|D_P^-(x)|} \geq l_1 \text{ and } \frac{|D_P^+(x) \cap Cl_t^\geq|}{|D_P^+(x)|} \geq l_2\}$

● High boundary region $P^{\beta l_1}(Cl_t)$ :

$P^{\beta l_1}(Cl_t) = Bn_P^{l_1 l_2}(Cl_t^\leq) \cap Cl_t = \{x \in Cl_t : \frac{|D_P^-(x) \cap Cl_t^\leq|}{|D_P^-(x)|} < l_1\}$

Particularly, there are just $\underline{P}^{l_1 l_2}(Cl_t)$ and $P^{\beta l_1}(Cl_t)$ for class $Cl_1$ and just $P_\beta^{l_2}(Cl_t)$ and $\underline{P}^{l_1 l_2}(Cl_t)$ for class $Cl_l$.

### 3.3. Class-based VP-DRSA model

Inuiguchi et al. [6] introduce the VP-DRSA model defined as follows: For any $P \subseteq C$, we say that $x \in U$ belongs to $Cl_t^\geq$ at precision level $l_2 \in (0,1]$, and $x \in U$ belongs to $Cl_t^\leq$ at precision level $l_1 \in (0,1]$. The concept of lower approximations at some precision levels $l_1$ and $l_2$ are formally presented as:

$\underline{P}^{l_2}(Cl_t^\geq) = \{x \in U : \frac{|D_P^-(x) \cap Cl_t^\geq|}{|D_P^-(x) \cap Cl_t^\geq| + |D_P^+(x) \cap Cl_{t-1}^\leq|} \geq l_2\}$ , $t = 1,...,l$ ;

$\underline{P}^{l_1}(Cl_t^\leq) = \{x \in U : \frac{|D_P^+(x) \cap Cl_t^\leq|}{|D_P^+(x) \cap Cl_t^\leq| + |D_P^-(x) \cap Cl_{t+1}^\geq|} \geq l_1\}$ , $t = 1,...,l$ .

Particularly, when $D_P^+(x) \subseteq Cl_t^\geq$, we have $D_P^+(x) \cap Cl_{t-1}^\leq = \emptyset$, and $l_2 = 1$. Accordingly, $\underline{P}^{l_2}(Cl_t^\geq)$ becomes DRSA lower approximation $\underline{P}(Cl_t^\geq)$. The same situation is happened in $\underline{P}^{l_1}(Cl_t^\leq)$.

The TRM with respect to the predefined class $Cl_t$ ($t = 2,...,l-1$) can then be presented as follows:

● Low boundary region $P_\beta^{l_2}(Cl_t)$ :

$P_\beta^{l_2}(Cl_t) = \{x \in Cl_t : \frac{|D_P^-(x) \cap Cl_t^\geq|}{|D_P^-(x) \cap Cl_t^\geq| + |D_P^+(x) \cap Cl_{t-1}^\leq|} < l_2\}$

● Precision classification region $\underline{P}^{l_1 l_2}(Cl_t)$ :

$$\underline{P}^{l_1 l_2}(Cl_t) = \{x \in Cl_t : \frac{|D_P^+(x) \cap Cl_t^\leq|}{|D_P^+(x) \cap Cl_t^\leq| + |D_P^-(x) \cap Cl_{t+1}^\geq|} \geq l_1$$

$$\text{and } \frac{|D_P^-(x) \cap Cl_t^\geq|}{|D_P^-(x) \cap Cl_t^\geq| + |D_P^+(x) \cap Cl_{t-1}^\leq|} \geq l_2\}$$

- High boundary region $P^{\beta l_1}(Cl_t)$:

$$P^{\beta l_1}(Cl_t) = \{x \in Cl_t : \frac{|D_P^+(x) \cap Cl_t^\leq|}{|D_P^+(x) \cap Cl_t^\leq| + |D_P^-(x) \cap Cl_{t+1}^\geq|} < l_1\}$$

Particularly, there are just $\underline{P}^{l_2}(Cl_t)$ and $P^{\beta l_1}(Cl_t)$ for class $Cl_1$ and just $P_\beta^{l_2}(Cl_t)$ and $\underline{P}^{l_1 l_2}(Cl_t)$ for class $Cl_l$.

### 3.4. A discussion

Let us remark the two extensions of classical DRSA model: VC-DRSA model and VP-DRSA model. We firstly take the consistency and precision in class union $Cl_t^\geq$ as example. In VC-DRSA model, consistency $\alpha$ can be defined by:

$$\alpha = \frac{|D_P^+(x) \cap Cl_t^\geq|}{|D_P^+(x)|} = \frac{|D_P^+(x) \cap Cl_t^\geq|}{|D_P^+(x) \cap Cl_t^\geq| + |D_P^+(x) \cap Cl_{t-1}^\leq|}.$$

If $\alpha = 1$ is satisfied, we have $|D_P^+(x) \cap Cl_{t-1}^\leq| = 0$. Then, we obtain $D_P^+(x) \subseteq Cl_t^\geq$, which abides by the strict dominance principle of classical DRSA model. In VP-DRSA model, precision $\beta$ can be defined by:

$$\beta = \frac{|D_P^-(x) \cap Cl_t^\geq|}{|D_P^-(x) \cap Cl_t^\geq| + |D_P^+(x) \cap Cl_{t-1}^\leq|}.$$

Similarly, if $\beta = 1$ is satisfied, we have $|D_P^+(x) \cap Cl_{t-1}^\leq| = 0$. Then, we obtain $D_P^+(x) \subseteq Cl_t^\geq$, which abide by the strict dominance principle of classical DRSA model. Comparing the definition of consistency $\alpha$ with that of precision $\beta$, the only difference is shown as followings:
(1) $\alpha$ is related to dominance cone $D_P^+(x)$;
(2) $\beta$ is related to dominance cone $D_P^+(x)$ and $D_P^-(x)$.

From the viewpoint of class-based rough approximation, we remark that the VP-DRSA model is focused on the Low and High boundary regions of TRM. More specifically, regarding the class $Cl_t$, precision degree $l_1$ is based on the investigation of objects $x \in P^\beta(Cl_t)$. Similarly, precision degree $l_2$ is derived from the exploitation of assignment information of boundary region: $x \in P_\beta(Cl_t)$. Therefore, for VP-DRSA, we have the following assertions:
(1) For each object $x \in Cl_t$, the real value $\beta_1$ represents: to what extent, object $x$ belongs to the High boundary region: $P^{\beta_1}(Cl_t)$, where,

$$\beta_1 = \frac{|D_P^+(x) \cap Cl_t^\leq|}{|D_P^+(x) \cap Cl_t^\leq| + |D_P^-(x) \cap Cl_{t+1}^\geq|}.$$

(2) For each object $x \in Cl_t$, the real value $\beta_2$ represents: to what extent, object $x$ belongs to the Low boundary region: $P_{\beta_2}(Cl_t)$, where,

$$\beta_2 = \frac{|D_P^-(x) \cap Cl_t^\geq|}{|D_P^-(x) \cap Cl_t^\geq| + |D_P^+(x) \cap Cl_{t-1}^\leq|}.$$

As such, the predefined levels $l_1$ and $l_2$ are used to control the precision degrees $\beta_1$ and $\beta_2$ in definitions of lower approximation, respectively.

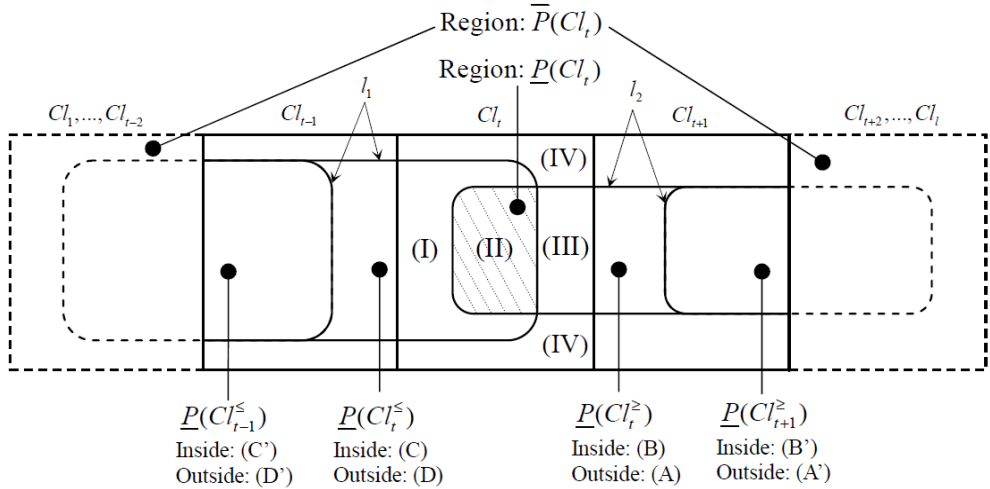

Figure 2. The illustration of constraint conditions preserving decision class $Cl_t$.

### 4. Class-based criteria reduction

Kusunoki and Inuiguchi [7] studied the definitions of class-based rough approximations and also provided the new concepts of class-based reducts. The definition is given as follows:

**Definition** For $P \subseteq C$ and $t \in T$, lower and upper approximations of decision class $Cl_t$ are defined by:

$\underline{P}(Cl_t) = \underline{P}(Cl_t^\geq) \cap \underline{P}(Cl_t^\leq)$ ; $\overline{P}(Cl_t) = \overline{P}(Cl_t^\geq) \cap \overline{P}(Cl_t^\leq)$ .

In this definition, $\underline{P}(Cl_t)$ is constrained by both conditions (B) and (C), which is also the precision classification region of TRM in classical DRSA model. And, $\overline{P}(Cl_t)$ is constrained by both conditions (D') and (A'). Please refer to the illustration of Fig. 2.

With this in mind, the following assertions are valid:

$Bn_P(Cl_t) = \overline{P}(Cl_t) - \underline{P}(Cl_t)$

$\underline{P}(Cl_t) = \{x \in Cl_t : D_P^+(x) \subseteq Cl_t^\geq \text{ and } D_P^-(x) \subseteq Cl_t^\leq\}$ ;

$\overline{P}(Cl_t) = \{x \in U : D_P^-(x) \cap Cl_t^\geq \neq \emptyset \text{ and } D_P^+(x) \cap Cl_t^\leq \neq \emptyset\}$ ;

$\overline{P}(Cl_t) = U - \underline{P}(Cl_{t-1}^\leq) - \underline{P}(Cl_{t+1}^\geq)$

$Bn_P(Cl_t) = U - \underline{P}(Cl_{t-1}^\leq) - \underline{P}(Cl_t) - \underline{P}(Cl_{t+1}^\geq)$

$\underline{P}(Cl_t) \subseteq Cl_t \subseteq \overline{P}(Cl_t)$

$\overline{P}(Cl_t) = Cl_t \cup Bn_P(Cl_t)$

$\overline{P}(Cl_t^\geq) = \bigcup_{k \geq t, k \in I} \overline{P}(Cl_k)$

$\overline{P}(Cl_t^\leq) = \bigcup_{k \leq t, k \in I} \overline{P}(Cl_k)$

$Bn_P(Cl_t) = Bn_P(Cl_t^\geq) \cup Bn_P(Cl_t^\leq)$

$\underline{P}(Cl_t) + \bigcup_{k \neq t, k \in I} \overline{P}(Cl_k) = U$

$\bigcup_{t \in I} \underline{P}(Cl_t) + \bigcup_{t \in I} Bn_P(Cl_t) = U$

For $P \subseteq C$, $\underline{P}(Cl_t) \subseteq \underline{C}(Cl_t)$ ; $\overline{P}(Cl_t) \supseteq \overline{C}(Cl_t)$ .

And also, the following assertion presented in [7] is actually not invalid:

$\underline{P}(Cl_t) = Cl_t - Bn_P(Cl_t)$ .

It can be revised as the following assertion for describing the relations among $Cl_t$, $\underline{P}(Cl_t)$ and $Bn_P(Cl_t)$:

$Bn_P(Cl_t) \cap Cl_t = Cl_t - \underline{P}(Cl_t)$

According to the proposed TRM, the class-based reducts can be defined as follows:

**Definition** (L-reduct):
If a minimal subset $P \subseteq C$ fulfills $\underline{P}(Cl_t) = \underline{C}(Cl_t)$ for $t = 1, ..., l$, this criteria subset is a Lower approximation reduct, denoted by L-reduct.

**Definition** (L$\beta$-reduct):
If a minimal subset $P \subseteq C$ fulfills $P_\beta(Cl_t) = C_\beta(Cl_t)$ for $t = 2, ..., l$, this criteria subset is an Low boundary reduct, denoted by L$\beta$-reduct.

**Definition** (H$\beta$-reduct):
If a minimal subset $P \subseteq C$ fulfills $P^\beta(Cl_t) = C^\beta(Cl_t)$ for $t = 1, ..., l-1$, this criteria subset is an High boundary reduct, denoted by H$\beta$-reduct.

**Proposition:**
If a criteria subset $P \subseteq C$ is the H$\beta$-reduct as well as the L$\beta$-reduct, we assert this subset $P$ is also the $L$-reduct.

*Proof.* It can be easily proved by using our proposed TRM of class-based rough approximation. □

## 5. Conclusion

Unlike the union-based definitions in previous DRSA models, this paper attempts to develop the class-based definitions of rough approximation. Based on the analysis of the partition in one singleton decision class, a new Three Region Model is proposed. In addition, we study the relationship of definitions between union-based rough approximations and class-based rough approximations. Several consequential properties are provided in this paper. Finally, we provided the new class-based reducts with assistance of the Three Region Model.